\documentclass[slac_one]{revtex4}
\usepackage{graphicx}
\usepackage{fancyhdr}
\begin{document}

\title{Observation of candidate exotic mesons containing heavy 
quarks with Belle} 

%
\author{S.-K. Choi (For Belle collaboration)}
\affiliation{Gyeongsang Nat. University, Jinju, Rep. of Korea}

\begin{abstract}
We report the observation of two resonance-like structures in the 
$\pi^+ \chi_{c1}$ invariant mass distribution near 4.1 GeV in exclusive 
$B\to K\pi^+\chi_{c1}$ decays.  A detailed Dalitz-plot analysis demonstrates 
that these structures cannot be produced by reflections from any known 
and possibly unknown resonances in the $K\pi$ channel. If these two peaks 
are produced by resonances in the $\pi^+\chi_{c1}$ channel, their minimal 
quark structure would have to be a $c\bar{c}u\bar{d}$ tetraquark 
arrangement, similar to that proposed for the $Z^+(4430)$ structure reported 
by Belle last year in the $\pi^+\psi^{'}$ mass distribution produced in  
$B\rightarrow K\pi^+\psi^{'}$ decays. In addition, we report 
some new measurements on the properties of the $X(3872)$ meson and the 
$1^{--}$ $Y$ states that are 
produced with initial state radiation. 
The analyses are based on a large data sample recorded at 
the $\Upsilon(4S)$ resonances and nearby continuum 
with the Belle detector at the KEKB asymmetric-energy $e^+e^-$ collider.
\end{abstract}

\maketitle

\thispagestyle{fancy}

\section{Introduction} 

The $B$-factories were built to study $CP$ violation in $B$-meson decays. They 
also serve as a plentiful source of $c \bar{c}-like $ quark pairs   
produced via the color-suppressed, 
but CKM-favored $b\rightarrow c\bar{c}s$ decay mode and by
$s$-channel $e^+e^-\rightarrow c\bar{c}$ annihilation via initial state radiation. 
A number of particles are observed, the so-called $XYZ$ mesons,  
that do not  fit into  any of the predicted but as yet unobserved 
charmonium states.
As a result, they are considered possible candidates for new, exotic types 
of particles such as multiquark states or $c\bar{c}$-gluon hybrids.
The former include either mesonic molecules ($c\bar{q}\bar{c} q$ ) 
or diquark-diantiquark ($cq\bar{c}\bar{q}$ ) tetraquark mesons.
We report on the recently observed $Z$ family of states that have 
non-zero charge, which means they can neither be $c\bar{c}$ nor hybrid 
states.  In addition new measurements related to the $X(3872)$ are mentioned,
and properties of the $1^{--}$ $Y$ family of states that are  produced 
via the initial state radiation process, including those of the
possible $s\bar{s}$ analog, the $Y_s(2175)$, are reported.

\section{Charged resonance-like $Z_{1}$ and $Z_{2} $ states}

\begin{figure*}[t]
\centering
\includegraphics[width=55mm]{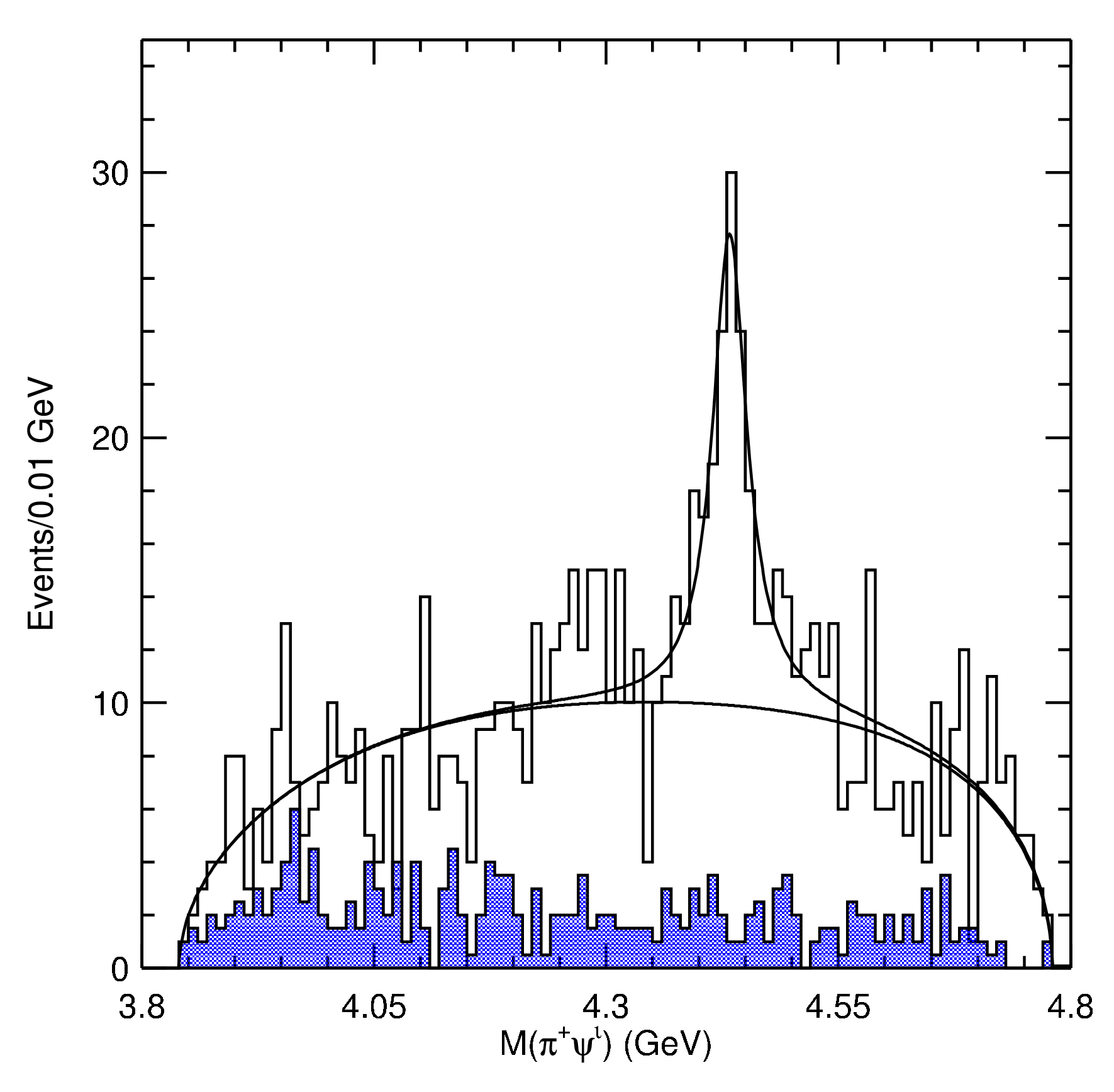} 
\caption{The $M(\pi^{+}\psi^{'})$ distribution produced in $B\rightarrow K \pi^{\pm} \psi^{'}$
decays. The $K^{*}$ veto was applied. The shaded histogram shows the combinatoric background
level inferred from side-band data.} 
\label{z4430}
\end{figure*}

In 2007 Belle reported the first charged charmonium-like 
state, denoted by $Z^{+}(4430)$,
in the $\pi^{+} \psi^{'}$ mass distribution produced in  
$B\rightarrow K \pi \psi^{'}$ decays~\cite{z4430-ref}.
After vetoing the peak regions of the $K^*(890)$ and $K^*_2 (1430)$ 
resonances in order to minimize their effects (the $K^{*}$ veto), 
the $\pi^+ \psi^{'}$ mass distribution (shown
in Fig.~\ref{z4430}) exhibits a strong enhancement near 4.43~GeV/$c^2$.
A fit with an $S$-wave Breit Wigner function for signal peak and
a phase-spacelike function for the non-peaking background
gives a mass and width of $M=4433\pm 4\pm 2 $~MeV/$c^2$ and 
$\Gamma = 45^{+18}_{-13}(stat)^{+30}_{-13}(syst)$~MeV/$c^2$ and a signal 
significance  of $6.5\sigma$. The product branching fraction is measured 
to be ${\it B} (B^{0} \rightarrow K^{-} Z^{+}(4430)) \times {\it B} 
(Z^{+}(4430) \rightarrow
 \pi^{+} \psi^{'}) = (4.1 \pm 1.0 \pm 1.4) \times 10^{-5}$.

This $Z(4430)^{+}$ observation motivated a search for other charged charmonium-like 
states in $\bar{B}^0 \rightarrow K^- \pi^+ c\bar{c}$ decays~\cite{z1z2-ref}.

\begin{figure*}[t]
\centering
\includegraphics[width=135mm]{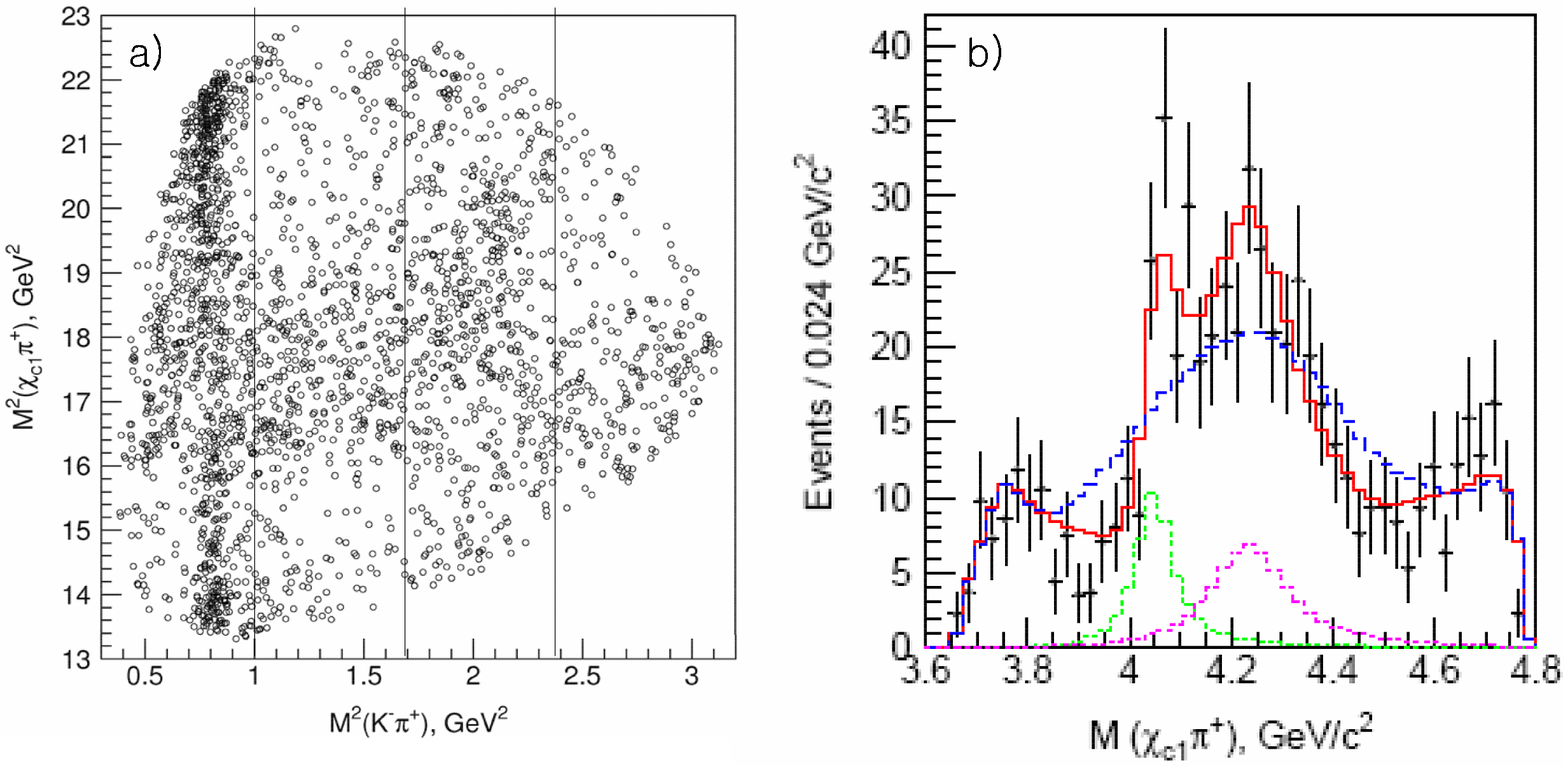} 
\caption{a) The $M^2 (K^- \pi^+)$ (horizontal) vs $M^2 (\pi^+ \chi_{c1})$ Dalitz plot 
distribution for the $B\rightarrow K \pi^{+} \chi_{c1}$ candidate events. 
b) The $M(\pi^{+} \chi_{c1})$ distribution projected from the second vertical slice 
in the range of 1.0 ${\rm GeV^{2}/}c^{4} < M^2 (K^{-} \pi^{+}) < $ 1.75 ${\rm GeV^2/}c^4 $  
in the Dalitz plot a), 
which is the most sensitive to $Z_1$ and $Z_2$.
The dots with error bars represent data, the solid (dashed) histogram is the 
Dalitz plot fit model with all known $K^*$ and two (without any) $\pi^+ \chi_{c1}$
resonances.} 
\label{z1z2}
\end{figure*}

The left panel of Fig.~\ref{z1z2} shows a Dalitz plot of $M^2 (K^+ \pi^-)$ 
(horizontal) versus $M^2 (\pi^+ \chi_{c1})$ (vertical) for 
$\bar{B}^0 \rightarrow K^-\pi^+ \chi_{c1}$ events.
Some clustering is evident in vertical bands at the mass of the $K^* (890)$ 
and $K^*_2 (1430)$, which correspond to $B\rightarrow K^{*} \chi_{c1}$ decays.
In addition, a clustering of events in a broad horizontal band is also apparent.
To characterize this signal clustering in the Dalitz plot, Belle did a likelihood fit
with a two-variable ($M^2 (K^- \pi^+)$
\& $M^2 (\pi^+ \chi_{c1})$) function that is a superposition of various 
interfering Breit-Wigner amplitudes that
includes all known $K^- \pi^+$ resonances plus $\pi^+ \chi_{c1}$ resonances. 
The binned fit to the Dalitz plot is performed using a very small bin size.

To display the fit results,
the Dalitz plot is divided into the four vertical slices shown in Fig.~\ref{z1z2}~a) and 
three horizontal slices.
The second vertical slice in Fig.~\ref{z1z2}~a) is the most sensitive to possible $Z^+$ signals;  
the first vertical slice has some $Z^+$ sensitivity.
The right plot in Fig.~\ref{z1z2}~b) shows the projected $\pi^+ \chi_{c1}$ mass 
distribution 
from the second vertical $K\pi $ mass slice, the region with most discrimination;
the histograms are the results of fits to the entire Dalitz plot that are described below.

The dashed histogram shows the results of a fit to the $\pi^+ \chi_{c1}$ structure  
with a fit function that includes all known $K\pi$ resonances,
but no $\pi^+ \chi_{c1}$ resonance. The fit CL is very poor, 
only $4 \times 10^{-12}$.
When we include a non-resonant $K \pi \chi_{c1}$ term and  an additional $K^*$ resonance 
with floating mass and width into the fit function, the fit CL 
(from a toy MC study) is still only $2 \times 10^{-4}$.
If we add one spin=0 $\pi \chi_{c1}$ resonance, the fit gives 
a mass and width of $M=4150 ^{+31}_{-10}$~MeV/$c^2$ and $\Gamma=352 ^{+99}_{-43}$~MeV/$c^2$  
with signal significance of 10.7 $\sigma$. However, the fit CL is only $0.1\%$.

If we instead use two $\pi \chi_{c1}$ resonances, denoted by $Z_1$ and $Z_2$,
the fit gives masses and widths of $M_{Z_{1}} = 4051\pm 14 ^{+20}_{-41} {\rm MeV/}c^{2}$,
$\Gamma_{Z_{1}} = 82 ^{+21}_{-17}(stat) ^{+47}_{-22}(syst)$~MeV/$c^2$, 
$M_{Z_{2}} = 4248^{+44} _{-29}(stat) ^{+180} _{-35}(syst) {\rm MeV/}c^{2}$ and 
$\Gamma_{Z_{2}} = 177^{+54}_{-39}(stat) ^{+316} _{-61}(syst)$~MeV/$c^2$, and a fit CL of 40$\%$.
The corresponding product branching fractions are
${\it B} (\bar{B}^{0} \rightarrow K^{-} Z^{+} _{1} ) \times {\it B} (Z^{+} _{1} \rightarrow 
\pi ^{+} \chi_{c1} ) = (3.1 ^{+1.5} _{-0.8}(stat) ^{+3.7} _{-1.6}(syst)) \times 10^{-5}$ 
and
${\it B} (\bar{B}^{0} \rightarrow K^{-} Z^{+} _{2} ) \times {\it B} (Z^{+} _{2} \rightarrow 
\pi ^{+} \chi_{c1} ) = (4.0 ^{+2.3} _{-0.9}(stat) ^{+19.7} _{-0.5}(syst)) \times 10^{-5}$.
These are comparable to those of other charmonium-like states as well as the $Z(4430)^+$.
The two resonances structure is favored over the one resonance assumption 
at the 5.7 $\sigma$ level.

We tried to fit with many different $K\pi$ resonance options including the
addition of new resonances with floating masses and widths.  However,
no model that only has dynamics in the $K\pi$ channel can fit the Dalitz plot.

\section{News on the $X(3872)$}

The threshold mass of the $D^0 \bar{D}^{*^0}$ system is $3871.8 \pm 0.4$~MeV/$c^2$, while
the PDG 2008 world average mass of the $X(3872)$ is $3871.9 \pm 0.5$~MeV/$c^2$.
This similarity in mass is a frequently cited characteristic of the $X(3872)$ and 
highly suggestive of  a $D \bar{D}^{*}$ molecule-like structure.
This motivated searches for $X(3872) \rightarrow D \bar{D}^{*}$ decays,
and peaks with masses that are slightly higher than 3872~MeV/$c^2$ were reported both by Belle 
($3875.2 \pm 0.7^{+0.3} _{-1.6} \pm 0.8 {\rm MeV/}c^2$)
and BaBar ($3875.1^{+0.7} _{-0.5} \pm 0.5 {\rm MeV/}c^2$).
Both goups also measured branching fractions that are 10 times that 
for the decay $ B \rightarrow K X(3872) $; $X(3872)\rightarrow J/\psi \pi^+ \pi^- $  .
This might be trouble for the $D\bar{D}^{*}$ mesonic-molecule interpretation~\cite{molecular} 
of the  $X(3872)$, for which the predicted BF ratio is more like 0.1.

Here we report results from a three-times-larger data sample of 657 million $B\bar{B}$ pairs and 
uses both $D^*$ decays to $D^0 \gamma$ and $D^0 \pi^0$ to see
 whether the threshold $D\bar{D}^*$ mass peak is really at a higher mass than 3872~MeV/$c^2$.
We also fit the signal with a more sophisticated form that uses
a Breit Wigner and the Flatte fitting function.
The Flatte function is essential if the resonance peak mass is  
below the $D\bar{D}^*$ mass threshold. Fits assuming that the peak is below or above
the threshold give the same signal significance of $8.8 \sigma$
and a fitted peak mass of the Breit Wigner function of
$3872.6^{+0.5} _{-0.4} \pm 0.4 {\rm MeV}/c^2$, which is consistent with
the current world averaged value for the $X(3872)$ mass from the $J/\psi \pi \pi$ channel. 
We also measure the product branching fraction 
for the $D\bar{D}^*$ mode to be $(0.73\pm 0.17 \pm 0.13) \times 10^{-4} $.

%


\begin{figure*}[t]
\centering
\includegraphics[width=115mm]{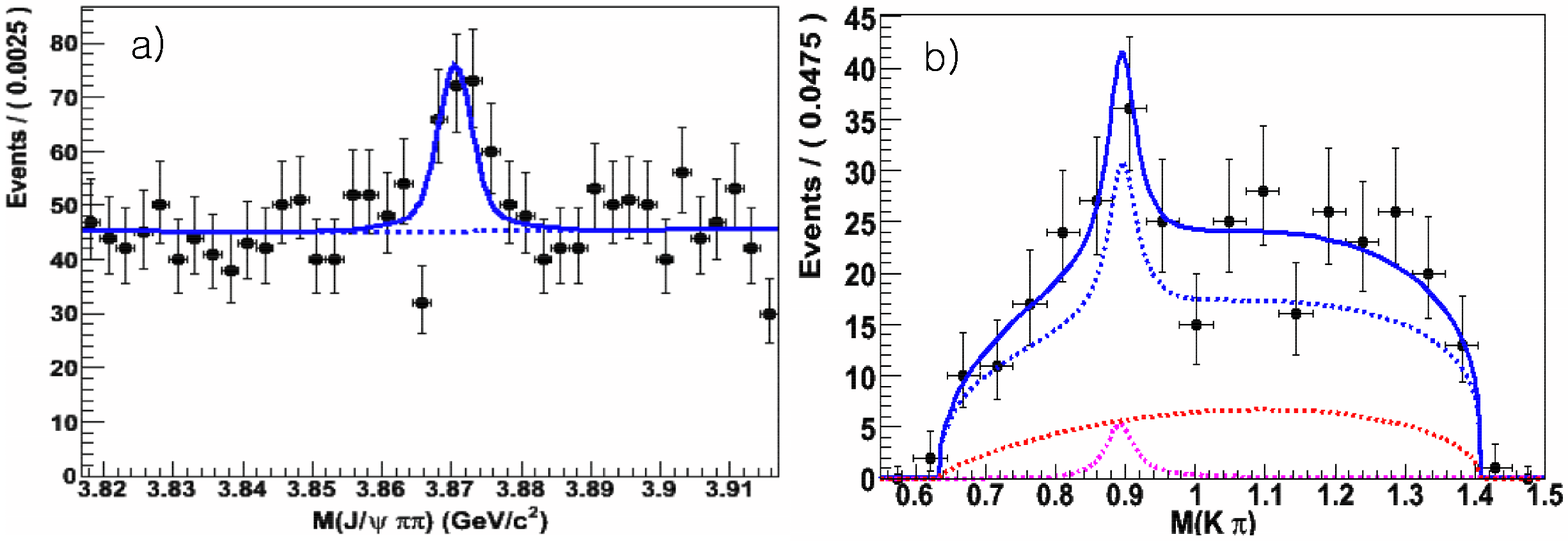}  
\caption{a) The $M(J/\psi \pi^+ \pi^- )$ distribution for the $X(3872)$ region produced in 
$B^0 \rightarrow J/\psi \pi^+ \pi^- K^+ \pi^-$ decays. b) The $K \pi$ 
mass distribution for $B \rightarrow X(3872) K \pi$ candidate events. 
The dash-dotted curve shows the non-resonant $K \pi$ component, the dotted
curve the $K^* (892)^{0}$ component and the dashed curve the background from sideband.    } 
\label{karimf}
\end{figure*}

In addition, we searched for the $X(3872)$ in the $J/\psi \pi^+ \pi^-$ mass distribution
for the $B^0 \rightarrow J/\psi \pi^+ \pi^-  K^+ \pi^-$ decays~\cite{karim}.
Figure~\ref{karimf}~a) shows a 5$\sigma$ significance $X(3872)$ signal 
in the $J/\psi \pi^+ \pi^- $ mass distribution 
from $B^0 \rightarrow J/\psi \pi^+ \pi^- K^+ \pi^-$ decays and Fig.~\ref{karimf}~b)
shows the $K\pi$ mass distribution for the $X(3872)$ mass region.
Here the non-resonant $K\pi$ component, shown by the dash-dotted curve, 
is much stronger than the 
resonant $K^* (892)^{0}$ portion shown by the dotted curve.
The measured product branching fraction is
${\it B} (B^0 \rightarrow X(3872) (K^+ \pi^-)_{non-res}) \times {\it B}(X(3872)
\rightarrow J/\psi \pi^+ \pi^- )= (8.1 \pm 2.0 ^{+1.1} _{-1.4} )\times10^{-6}$ 
at the $90\%$ CL and ${\it B} (B^0 \rightarrow X(3872) (K^{*}(890)^0) \times {\it B}(X(3872)
\rightarrow J/\psi \pi^+ \pi^- )< 3.4 \times10^{-6}$.
This feature is in contrast to corresponding values for 
conventional charmonium ($c\bar{c}$) states such as 
$J/\psi, \psi^{'}$ and $\chi_{c1}$,
where the $ B \rightarrow (c\bar{c}) K^{*}(890)^0$ decay is much larger 
than non-resonant $ B \rightarrow (c\bar{c})K^+ \pi^- $ contributions.

We also report the first statistically significant observation of 
$ B^0 \rightarrow X(3872) K^0 _{S} $. The ratio of branching fractions is measured  
to be $\frac{B^0 \rightarrow X(3872) K^0 _{S}}
{B^+ \rightarrow X(3872) K^+)}=0.82\pm0.22\pm0.05$, which is consistent with unity.
The mass difference between the $X(3872)$ produced in $B^+$ and $B^0$ decays
is measured to be $\Delta M = M_{XK^+} - M_{XK^0}= 0.18 \pm 0.89 \pm 0.26~{\rm MeV/}c^2$,
which is cosisitent with 0.

\section{$Y$ states produced via ISR}

The $Y$ states are $1^{--}$ states produced via initial state radiation (ISR). 
Both the  $Y(4260)$~\cite{y4260-babar} seen in $J\psi \pi^+ \pi^-$ masses 
and the $Y(4325)$~\cite{y4325-babar} seen in $\psi^{'} \pi^+ \pi^-$ masses were 
discovered in $e^+e^- \rightarrow \gamma_{ISR} Y $ processes by BaBar. 
They are not the same state.

Belle confirmed both states with the same decay modes as BaBar's also using ISR.
However, Belle observed one more peak in the $\pi\pi J\psi$ channel (denoted by $Y(4040)$) 
with a mass of $M=4008 \pm 40 ^{+114} _{-28} $~MeV/$c^2$ and 
the width of
$\Gamma = 226 \pm 44 \pm 87$~MeV/$c^2$ with a significance of $7.4\sigma$. 
This state's peak mass is consistent with the
$\psi(4040)$ charmonium state in mass, but its measured width is
much larger~\cite{y4260-belle}.
Belle also observed an additional peak in the $\pi\pi \psi^{'}$
channel (designated as $Y(4660)$)
with mass  $M=4664\pm 11 \pm 5$~MeV/$c^2$ \& width
$\Gamma = 48 \pm 15 \pm 10$~MeV/$c^2$, and a significance of $5.8 \sigma$~\cite{y4325-belle}.
The $Y(4660)$ also does not fit neatly to any available charmonium assignments.
Among these results, one interesting feature is that the 
$\pi\pi \psi^{'}$ peak masses
are different from those seen in the $\pi \pi J/\psi$ channel.

\begin{figure*}[t]
\centering
\includegraphics[width=65mm]{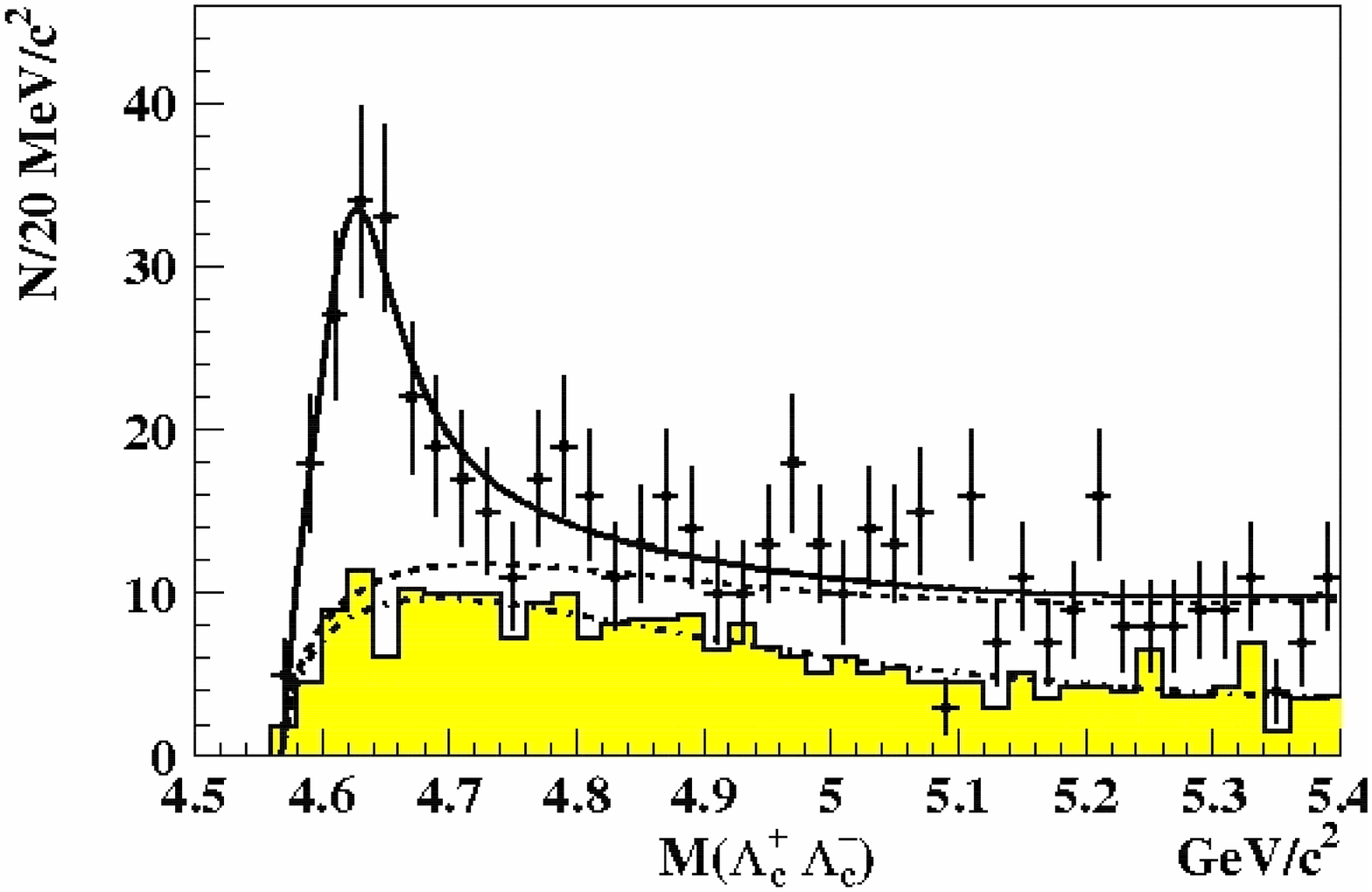} 
\caption{The $M(\Lambda^{+} _{c} \Lambda^{-} _{c}) $ distribution with 
$\bar{p}$ tag in the $e^+ e^-
\rightarrow \Lambda^{+} _{c} \Lambda^{-} _{c} $ cross section via initial state 
radition. 
The solid curve shows the fit result of the signal. 
The dashed curve shows that of the threshold function. The 
combinatorial background parametrization is shown by the dashed-dotted curve. 
} \label{lamlamb}
\end{figure*}

In addition, Belle observed a new near-threshold enhancement in the 
$e^+e^- \rightarrow \Lambda^{+} _{c} \Lambda^{-} _{c}$ cross section 
using initial-state radiation~\cite{galina}.
Figure~\ref{lamlamb} shows a clear peak near the
$\Lambda^{+} _{c} \Lambda^{-} _{c}$ invariant mass threshold.
Here only one $\Lambda_c$ is reconstructed and the other is inferred kinematics when 
only the decay antiproton is detected.
The signal function is a sum of a relativistic $S$-wave Breit-Wigner function
plus a threshold function to account for non-resonant production. 
The fit result gives a mass and width of
$M=4634^{+8} _{-7} (stat)^{+5}_{-8}(syst)~{\rm MeV/}c^2$ and $\Gamma = 92^{+40} _{-24} (stat)
^{+10} _{-21}(syst)~{\rm MeV/}c^2$, respectively.
These values are consistent with the $Y(4660)$ mass and width. 
This state can be interpreted as a dibaryon threshold effect such as the one
first reported by the BES experiment in the $M(p\bar{p})$ from the 
decay $J/\psi \rightarrow \gamma p \bar{p}$~\cite{BES-ppb}. 
However, the possibility of its being the $5^3 S_{1}$ charmonium state is 
not excluded.

\begin{figure*}[t]
\centering
\includegraphics[width=135mm]{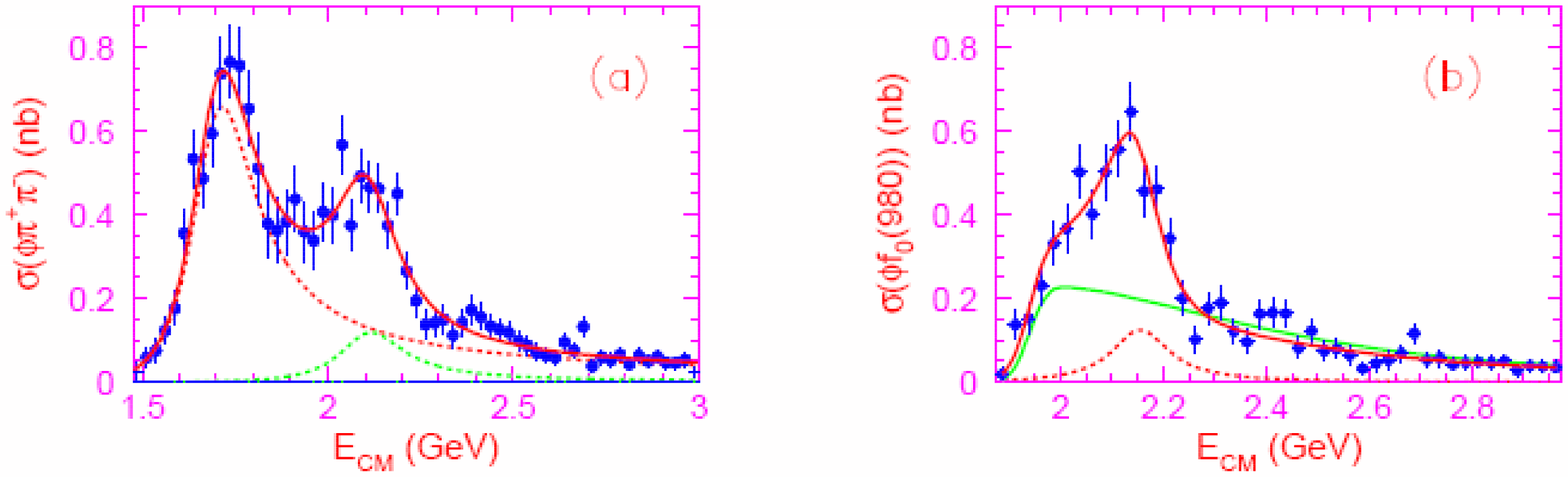}
\caption{a) The $e^+ e^- \rightarrow \phi \pi^+ \pi^-$ cross section
distribution fit with two coherent BW functions. b) The 
$e^+ e^- \rightarrow \phi f_0 (980)$ cross section distribution 
fit with one resonance and a coherent nonresonant term .} 
\label{y2175}
\end{figure*}

An interesting question is whether or not there exist any $XYZ$ counterparts
containing $s\bar{s}$ or $b\bar{b}$ quarks. Here we mention a 
possible candidate in $s\bar{s}$ sector. 
The $Y(2175)$ is a $1^{--}$ state first reported by BaBar 
in the $f_0 \phi$ mass produced in $e^+ e^- \rightarrow \gamma_{ISR} f_0 \phi$
ISR events; the mass and width are 
$M=2170\pm 10\pm 15~{\rm MeV/}c^2$ and 
$\Gamma = 58\pm 16 \pm 20~{\rm MeV/}c^2$~\cite{y2175-babar}.
Its producton and decay properties are similar as that of $Y(4260)$
and this has stimulated speculation that it may be its
$s\bar{s}$ counterpart, the  $Y_s (2175)$.
This state was confirmed by the BES experiment in $J/\psi\rightarrow \eta\phi f_0(980)$ decays
with a mass and width that agree well with the BaBar measurements~\cite{y2175-bes}.

We also measure the $\phi \pi^+ \pi^-$ and $\phi f_0$ cross section 
produced from the  $e^+ e^- \rightarrow \gamma_{ISR} \phi \pi^+ \pi^-$ and
$e^+ e^- \rightarrow \gamma_{ISR}\phi f_0$ initial state radiation events  
for the center-of-mass energy ranging from 1.5 to 3.0~GeV~\cite{y2175-belle}.
Figure~\ref{y2175} shows corresponding cross section distributions.

A fit with two coherent BW functions is applied to the $\phi\pi^+ \pi^-$ 
cross section distribution; a fit with one BW function interfering with a non-resonant 
continuum function is applied to the $\phi f_0(980)$ distribution.
By averaging the fit results near the mass of the $Y(2175)$, we obtain
a mass and width of
$2133^{+69} _{-115} {\rm MeV/}c^2$ and $169^{+105} _{-92} {\rm MeV/}c^2$, respectively.
The large uncertainty includes both statistical and systematic errors, where the systematic 
error includes 
the differnce of measured masses from $\phi \pi^+ \pi^-$ and $\phi f_0$ sample as well.
This mass agrees with both the BaBar and BES measurements, albeit with a somewhat larger width.
The fit to the low-mass peak in Fig.~\ref{y2175}~a) gives a mass and width of 
$M=1687\pm 21 {\rm MeV/}c^2$ and $\Gamma=212 \pm 29 {\rm MeV/}c^2$, respectively,
indicating that this is the $\phi(1680)$ radially excited $s\bar{s}$ state.
The possibility remains that the $Y(2175)$ could be just a higher radially excited
$s\bar{s}$ state.

\section{Summary}

We observe two more charmonium-like states, 
$Z^+ _1$ and $Z^+ _2$, with non-zero charge where both decay into $\chi_{c1}\pi^{+}$.
Their quark content is $c\bar{c}u\bar{d}$.
The $X(3872)$ decaying into $\pi^+ \pi^- J/\psi$ has been well established, 
while its underlying nature has not been conclusively identified. 
The favored interpretations are either a mesonic molecule
or a diquark-diantiquark tetraquark meson.
Some additional properties of the $X(3872)$ are measured.
The mass of the $X(3872)$ decaying into the $D\bar{D}^{0}$ final state  
is consistent with that from $\pi^+ \pi^- J/\psi$ within errors.
Another feature is that the non-resonant $K^+ \pi^{-}$ component in 
$B^0 \rightarrow X(3872) K^+ \pi^- ; X(3872)\rightarrow J/\psi \pi^+ \pi^-$
decay dominates over the resonant $K^* (892)^{0}$ component.
This is not similar to the behaviour usually seen for conventional charmonium states. 
Among several $1^{--}$ $Y$ states produced via initial state radiation, 
we observe the $Y(2175)$ as a possible $s\bar{s}$ analogue to the $Y(4260)$ which, in turn, is
a prime candidate for a  $c\bar{c}$-gluon hybrid. The $Y(4660)$ is also seen
via its $\Lambda^{+} _{c} \Lambda^{-} _{c}$ decay mode.  


\end{document}